# Dispersive detection of a charge qubit with a broadband high-impedance quantum-Hall plasmon resonator

Chaojing Lin[1,2*], Kosei Teshima[1], Takafumi Akiho[3], Koji Muraki[3], and Toshimasa Fujisawa[1]

[1]Department of Physics, Institute of Science Tokyo, 2-12-1 Ookayama, Meguro, Tokyo 152-8551, Japan.

[2]JST, PRESTO, 4-1-8 Honcho, Kawaguchi, Saitama 332-0012, Japan.

[3]Basic Research Laboratories, NTT, Inc., 3-1 Morinosato-Wakamiya, Atsugi, Kanagawa 243-0198, Japan.

*Corresponding author: Chaojing Lin (lin.c.7c98@m.isct.ac.jp)

**Abstract**

Cavity quantum electrodynamics (cQED) provides strong light-matter interactions that can be used for manipulating and detecting quantum states. The interaction can be enhanced by increasing the resonator's impedance, while approaching the quantum impedance ($h/e^2$) remains challenging. Edge plasmons emergent as chiral bosonic modes in the quantum Hall channels provide high quantized impedance of $h/\nu e^2$ that can exceed 10 kΩ for the Landau-level filling factor $\nu \leq 2$, well beyond the impedance of free space. Here, we apply such a high-impedance plasmon mode in a quantum-Hall plasmon resonator to demonstrate dispersive detection of a nearby charge qubit formed in a double quantum dot. The phase shift in microwave transmission through the plasmon resonator follows the dispersive shift associated with the qubit state in agreement with the cQED theory. The high impedance allows us to perform dispersive detection of qubit spectroscopy with a plasmon resonator having a broad bandwidth. Leveraging these topological edge modes, our results establish two-dimensional topological insulators as a new platform of cQED.

**Introduction**

Quantum Hall (QH) states are well-understood two-dimensional topological insulators showing highly quantized Hall conductance in the units of $e^2/h$.[1-2] Although the non-interacting picture is conveniently used to describe the edge transport, the low-energy excitation should be described by collective plasmon modes, particularly when highly nonequilibrium or high-frequency transport is concerned[3-6]. Interestingly, the impedance of the plasmon mode is also quantized as $h/\nu e^2$ at the Landau-level filling factor $\nu$.[5,7] Thus, the plasmons with this fixed impedance should show perfect impedance matching over the whole channel even in the presence of non-uniform electrostatic environment. Because the plasmon mode in the QH regime is theoretically constructed from electrons in terms of the bosonization scheme, the interaction-induced emergent quasiparticles should follow bosonic characteristics.[8,9] Actually, a recent experiment demonstrates plasmon-assisted tunneling in a double quantum dot (DQD) coupled to a plasmon resonator, where integer multiples of plasmons are absorbed by or emitted from the DQD.[10] This implies possible applications for cavity quantum electrodynamics (cQED)[11,12]. Since the vacuum fluctuation of the electric potential in the resonator increases with the resonator's impedance, strong cQED effects are expected with high impedance ($Z$ = 13 kΩ at ν = 2), as compared to conventional 3D cavities ($Z$ = 377 Ω) and sophisticated Josephson-junction arrays (a few kΩ) for photons[13-17]. While strong, ultra-strong, and deep strong coupling regimes are expected in plasmonic cQED systems[10,12], their experimental realizations remain challenging. One of the applications of cQED is the dispersive detection of a qubit, where the transmission amplitude and phase of photons (or plasmons in our case) are sensitive to the qubit state. Because the signal is enhanced by the quality factor $Q$ of the resonator or cavity, high-$Q$ resonators are usually appreciated at the cost of long measurement time due to their narrow bands[18-20]. However, one can choose a high-$Z$ and low-$Q$ resonator to achieve broadband dispersive detection.

Here, we study the plasmon transmission characteristics of a plasmonic resonator with a high $Z$ of 13 kΩ and a low $Q$ of 4 coupled to a semiconductor charge qubit in a DQD. A clear dispersive response is observed by changing the qubit parameters, i.e., the energy detuning and tunneling coupling of the DQD. Moreover, the broadband resonator allows us to demonstrate qubit spectroscopy over a wide energy range. These characteristics agree well with the dispersive shift calculated by the standard Jaynes-Cummings model. The demonstrated dispersive shift is obtained without significantly populating the qubit excited state. Some outlooks for strong coupling regimes are discussed.

**Results**

**Roles of $Z$ and $Q$ in dispersive shift**

We consider a cQED system schematically shown in Fig. 1a, where the charge qubit in the DQD is coupled to the closed edge channel forming a plasmon resonator with the perimeter of one plasmon wavelength. We measure the plasmon transmission coefficient $S_{21}$ by sending a microwave at probe frequency $f_\mathrm{p}$ from port 1 and receiving the signal at port 2 to investigate the qubit-resonator system. For a broadband resonator with low $Q$, we can vary $f_\mathrm{p}$ around the bare resonant frequency $f_\mathrm{r}$ of the resonator over the wide frequency width of about $\frac{f_\mathrm{r}}{Q}$. Here, the amplitude of electric potential for a given plasmon power in the resonator is enhanced by the large impedance $Z$. This potential can induce qubit transition near the resonant condition ($f_\mathrm{r} = f_\mathrm{q}$), where $f_\mathrm{q}$ is the transition frequency of the qubit, and dispersive shift is expected at large qubit-plasmon detuning $\Delta_{\mathrm{q-r}} = hf_\mathrm{q} - hf_\mathrm{r}$ in energy. The dynamics is characterized by the coupling strength $g = f_\mathrm{r}\eta\sin\theta_\mathrm{q}\sqrt{\frac{Ze^2}{h}}$ in frequency with a dimensionless factor $\eta\sin\theta_\mathrm{q}$ (<1) described below. Under the dispersive regime with large $\Delta_{\mathrm{q-r}} \gg hg$, the plasmon resonant frequency $f_\mathrm{r}' = f_\mathrm{r} + \Delta f_\mathrm{r}$ is shifted by $\Delta f_\mathrm{r} \cong -hg^2/\Delta_{\mathrm{q-r}}$, if the qubit is prepared in the ground state. This dispersive shift can be measured by the phase $\phi = \arg(S_{21}) \cong 2Q\Delta f_\mathrm{r}/f_\mathrm{r}$ of $S_{21}$ at $f_\mathrm{p} \sim f_\mathrm{r}$. Because $\phi$ is proportional to $QZ$ from the above formulas, a reasonable signal amplitude is expected with a high-$Z$ and low-$Q$ resonator.

**Device**

The device was fabricated in a GaAs/AlGaAs heterostructure (see Methods: Device Fabrication) hosting a two-dimensional electron gas (2DEG) 100 nm below the surface with an electron density $n_e$ = $1.8 \times 10^{11}$ cm$^{-2}$ and as-grown mobility μ ~ 460 m$^2$/Vs, as illustrated in Figs. 1a and 2a. The QH characteristics of the device are presented in Supplementary Note 1. All measurements were performed at the base lattice temperature of $T_B \simeq$ 100 mK. Under a perpendicular magnetic field $B$, the plasmon resonator (the red line with the arrows) is formed by closing the upper and lower constrictions by applying negative voltages on the gates labeled $G_{NU}$ and $G_{NL}$. The ports 1 and 2 are connected to the arc-shaped metal electrode $G_{inj}$ and $G_{det}$ (light pink), respectively, which are capacitively coupled to the resonator. Under applying dc voltage $V_{Gres}$ = -0.5 V on $G_{inj}$ and $G_{det}$, application of a microwave voltage to $G_{inj}$ displaces charge in the nearby 2DEG and excites plasmons (charge density waves) in the resonator. The plasmons travel clockwise along the perimeter $L$ = 113 μm of the disk-shaped 2DEG region (light orange) and are probed with $G_{det}$. Corresponding $S_{21}$ is measured with a vector network analyzer or a homodyne detection of in-phase and quadrature-phase components, as shown in Figs. 1b and 2b (see Methods: Spectrum Measurements and Supplementary Note 2 for the modeling).

The plasmon modes are confirmed by measuring $S_{21}$ at various $B$ and $f_p$. For convenience, we evaluate the relative change $\Delta|S_{21}| = |S_{21}/S_{21}^0|$ from the reference $S_{21}^0$ measured when the resonator is disabled by opening the lower constrictions $G_{NL}$ to connect the edge channel to the lower ohmic contact S in Fig. 1a (see Supplementary Note 3 for details). The representative data in Fig. 1c shows a distinct peak labeled $n_r$ = 1 and a faint peak labeled $n_r$ = 2, both of which follow the clear $1/B$ dependences (the solid lines). These dependences agree well with the edge magnetoplasmon modes.[7,21] Excess charge density ρ in the channel induces voltage $V_{ch} = \rho/C_\Sigma$ for the effective channel capacitance $C_\Sigma$ ($\cong$ 0.28 nF/m), and resulting Hall current $I = \sigma_{xy} V_{ch}$ flows along the channel at the plasmon velocity $v_p = \sigma_{xy}/C_\Sigma$ for the Hall conductance $\sigma_{xy}$. For the perimeter $L$, the resonant frequencies appear at $f_r = n_r v_p/L$ with integer $n_r$. Therefore, $f_r$ can be changed by tuning $\sigma_{xy}$ with $B$, as shown in Fig. 1c, and by tuning $C_\Sigma$ with the gate voltages $V_{Gres}$, as shown in Supplementary Fig. 3c. The overall $f_r \propto 1/B$ dependence in Fig. 1c can be understood with the classical Hall conductivity $\sigma_{xy} = en_e/B$, and the pronounced peaks with narrow width are seen at around the integer filling factors ν = 1, 2 and 3 (see Supplementary Note 4 for details). In the following experiments, we focus on the robust QH state at ν $\cong$ 2 ($B$ = 3.6 T). The frequency dependences of the amplitude $\Delta|S_{21}|$ in Fig. 1d and the phase shift $\Delta\phi$ = arg($S_{21}/S_{21}^0$) in Fig. 1e show the resonant frequency $f_r$ = 2.45 GHz, plasmon decay rate κ $\cong$ 0.6 GHz, and $Q = f_r/\kappa$ ~ 4 for $n_r$ = 1. We shall show below that dispersive detection is successful for this low-$Q$ and high-$Z$ resonator.

We now activate the DQD with quantum dots QD1 and QD2, which are located close to the resonator. Instead of measuring the current through the DQD (see Supplementary Note 5), we probe the DQD by measuring $S_{21}$. Tunneling between the two dots and between the dot and the lead induce finite change in $S_{21}$ when the tunneling rate is comparable to or greater than the plasmon frequency $f_p$, as presented in Supplementary Note 6. In Fig. 2c, we plot $\Delta\phi = \arg(S_{21}/S_{21}^0)$ measured at $f_p$ = 2.5 GHz as a function of the two gate voltages $V_{g1}$ and $V_{g2}$, which tune the electron numbers ($N_1$, $N_2$) in QD1 and QD2, respectively, under appropriate gate voltage $V_c$ for reasonable tunnel coupling. The negative peaks (blue for $\Delta\phi < 0$) elongated to the upper-right direction can be understood as due to interdot charge transitions between ($N_1$, $N_2$+1) and ($N_1$+1, $N_2$). The hexagonal charge stability diagram for ($N_1$, $N_2$) is outlined by the black dashed lines. An excess electron occupies QD1 and QD2 on the left and right sides of the transition signal, respectively, as shown in the inset to Fig. 2c. The energy difference ε between the localized states in QD1 and QD2 can be swept by changing $V_{g1}$ and $V_{g2}$ in the opposite directions as shown by the red arrow labeled ε. The average qubit energy δ, measured from the chemical potential of the leads, can be changed by sweeping the gate voltages in the orthogonal direction labeled δ. Around the center of the peak (δ = 0) in the δ direction, the qubit is effectively isolated from the leads due to the Coulomb blockade with the inter-dot charging energy of $\Delta_{12}$ ~74 μeV (see Supplementary Note 5). The dimensionless coupling strength η ~ $d\varepsilon/d(eV_{ch})$ measures how much energy in ε is modulated by the plasmon potential $eV_{ch}$. The appearance of the signal suggests finite η ($\cong$ 0.032 estimated from the following analysis), which is not optimized yet but sufficient to study the dispersive shift.

**Dispersive shift**

The qubit excitation energy $hf_q = \sqrt{\varepsilon^2 + 4t_c^2}$ is given by the tunnel coupling $t_c$ and the energy difference ε.[22] Therefore, $hf_q$ takes the minimum value of $2t_c$ at the center of the peak (ε = 0). Based on the dispersive shift $\Delta f_r \cong -hg^2/\Delta_{q-r}$ discussed above, the negative phase shift ($\Delta\phi < 0$) is associated with the negative frequency shift ($\Delta f_r < 0$) under the positive qubit-resonator detuning $\Delta_{q-r} = hf_q - hf_r > 0$. The simple negative peaks seen in Fig. 2c suggest that $2t_c > hf_r$. We can reach zero detuning ($\Delta_{q-r} = 0$) by reducing the interdot tunnel coupling $t_c$ via gate voltage $V_c$, as shown in Fig. 2d. While a single negative peak in $\Delta\phi$ is seen in panel (i) at $V_c$ = −413 mV, the single negative peak splits into two with a 'w' shaped profile in (ii)-(iv). $\Delta\phi$ at ε = 0 turns out to be positive in (iv) at $V_c$ = −415 mV, which indicates that $2t_c$ is reduced below $hf_r$. Zero detuning ($\Delta_{q-r} = 0$) is reached at the boundary of positive and negative $\Delta\phi$ regions.

We calculate the dispersive shift by using the standard cQED analysis[20], as summarized in Methods: Quantum Master Equation. The Jaynes-Cummings Hamiltonian describes the coupling between the resonator and the qubit with coupling strength $g = f_r \eta \sin\theta_q \sqrt{\frac{Ze^2}{h}}$. Here, η (< 1) is the electrostatic coupling coefficient, and $\sin\theta_q = \frac{2t_c}{\sqrt{\varepsilon^2 + 4t_c^2}}$ describes the relative amplitude of the transition dipole moment of the qubit[22]. Because $\sin\theta_q$ depends on ε and is maximized to $\sin\theta_q$ = 1 at ε = 0, we shall describe $g = g_c \sin\theta_q$ with maximal coupling $g_c = f_r \eta \sqrt{\frac{Ze^2}{h}}$ at ε = 0. The coupling to the environment is considered by using the Markovian quantum master equation with resonator decay rate κ, qubit relaxation rate $\gamma_1$, and qubit dephasing rate $\gamma_\phi$. Here, we consider the thermal plasmon number $n_{\mathrm{pl,th}} = \left(e^{\frac{hf_r}{k_B T_B}} - 1\right)^{-1} \simeq 0.45$ in the resonator. We assumed that the qubit is coupled to the phonon environment with the thermal phonon number $n_{\mathrm{ph,th}} = \left(e^{\frac{hf_q}{k_B T_B}} - 1\right)^{-1}$ ranging from ~ 0 at ε >> $k_B T_B$ to 0.48 at ε = 0 and $2t_c/h$ = 2.4 GHz. The input-output theory for the probe frequency $f_p$ is used to calculate $S_{21}$ (see Methods: Input-Output Theory for details).

We analyze the dispersive shift, i.e., how Δϕ changes with ε, for two representative data shown in the bottom panels (the circles) of Fig. 3a for $2t_c < hf_r$ and Fig. 3b for $2t_c > hf_r$. The simulations (the solid lines) reproduce the experimental data very nicely by choosing the parameters $t_c, \gamma_1, \gamma_\phi, g_c$ (shown in the figure caption). Because $\gamma_1$ and $\gamma_\phi$ cannot be separated with high accuracy, we evaluate the decoherence rate $\gamma = \gamma_1/2 + \gamma_\phi$ in the following. The fitting suggests the maximal coupling $g_c \cong 55$ MHz for both cases, while γ depends on $t_c$ (γ ≃ 0.34 GHz at $\frac{t_c}{h} \simeq 1.2$ GHz in Fig. 3a and 0.07 GHz at $\frac{t_c}{h} \simeq 1.43$ GHz in Fig. 3b). Whereas our particular device was in the weak coupling regime ($g_c < \gamma, \kappa$), the experiment was successful in demonstrating the dispersive detection of the qubit. The expected coupling dependency of $g_c = f_r \eta \sqrt{\frac{Ze^2}{h}}$ implies a relatively small η ≅ 0.032, which can be improved, as discussed below. The upper panels of Figs. 3a and 3b show the estimated eigen-energies with $g_c = 55$ MHz (the solid lines) and the uncoupled bare energies with $g_c = 0$ (the dashed lines). The evolution of Δϕ reflects the eigenstates of the system, and the conditions for Δϕ = 0 (the black circles in Fig. 3a) coincide with the resonance at $\Delta_{q-r} = hf_q - hf_r = 0$ (the crossing points of the dashed lines at ε = ±$\varepsilon_0$).

We repeated similar measurements at various $V_c$, as summarized in Fig. 3c, with measurements at various $f_p$ shown in Supplementary Note 7. The single negative peak observed at $V_c$ > -413 mV changes to the 'w' shaped profile at $V_c \leq$ -413 mV. The feature is reproduced well with the simulations (the red solid lines). The fitting parameters $t_c$ and γ are plotted as a function of $V_c$ in Figs. 3d and 3e, respectively. The qubit parameter $2t_c$ can be tuned smoothly across $hf_r$ (the dashed line in Fig. 3d) with $V_c$. γ in Fig. 3e decreases with increasing $V_c$ and thus with increasing $t_c$. This dependency implies that low-frequency fluctuation in ε is the dominant decoherence source[23-25], and thus intrinsic γ must be lower. Furthermore, the obtained $2t_c$ is plotted as a function of the resonant condition $\varepsilon_0$ in Fig. 3f. The dependence is consistent with the qubit energy $hf_q = \sqrt{\varepsilon_0^2 + 4t_c^2}$ at $f_q$ = 2.5 GHz (the red solid line). In this way, the plasmon resonator successfully measures the qubit characteristics.

**Qubit depolarization**

When the input RF power $P$ is increased, the average plasmon number $n_{\text{pl}} = \langle a^+ a \rangle = n_{\text{pl,th}} + n_{\text{pl,drv}}$ increases with the driven plasmon number $n_{\text{pl,drv}} \simeq \left(\frac{2eV_\text{d}}{h\kappa}\right)^2$ from the thermal one $n_{\text{pl,th}} \simeq$ 0.45, where $V_\text{d}$ is the drive voltage on the channel (See Methods: Quantum Master Equation). Under strong plasmon drive, the electron in the qubit is no longer in the ground state (Fig. 4a). As a result, the occupation probability $P_\text{g}$ in the ground state decreases, and $P_\text{e}$ in the excited state increases with increasing $P$, as shown in Fig. 4b, calculated for the experimental conditions for Fig. 3a. The qubit polarization $P_z = P_e - P_g = \langle \sigma_z \rangle$ is fully depolarized ($P_z = 0$) in the large limit of $V_\text{d}$ ($n_{\text{pl,drv}} \gg 1$), as shown in the upper scale of Fig. 4b. This depolarization reduced by half at around $V_\text{d} \simeq 10$ μV (blue circle), which is used to evaluate $V_\text{d}$ and $n_{\text{pl}}$ in our resonator.

For this purpose, we choose the condition close to the resonance at ε = 0 and $\frac{2t_c}{h} \simeq 2.4$ GHz close to $f_\text{r}$ = 2.5 GHz, which is identical to that in Fig. 3a. As $P$ is increased, the phase $\Delta\phi$ and amplitude $\Delta A$ at ε = 0 changes, as shown in Figs. 4c and 4d (see Supplementary Note 8 for details). Here, $\Delta\phi$ and $\Delta A$ are normalized by the average values taken in the low-power region ($P$ < -100 dBm). Both phase and amplitude decrease with increasing $P$ and are halved at $P \cong -85$ dBm for $\Delta\phi$ and $P \cong -76$ dBm for $\Delta A$. These dependences are reproduced with simulations of $\Delta\phi$ and $\Delta A$ by assuming a different proportionality factor $k$ in the relation $V_\text{d} = k\sqrt{P}$ (see Methods: Input-Output theory), as shown by the blue line for the $V_\text{d}$ scale in each figure. The different $k$ might be regarded as uncertainty of the estimate, because the theory suggests the same dependency for non-zero $\Delta_{\text{q-p}}$. Nevertheless, the high-$Z$ and low-$Q$ resonator is sensitive to $V_\text{d}$ and the corresponding plasmon number $n_{\text{pl,drv}}$ down to 10 or even lower. This analysis ensures that the dispersive detection in Fig. 3 with $P$ = -87.5 dBm, marked by the vertical gray dashed line in Fig. 4c and 4d, was performed without significantly populating the qubit excited state.

**Prospects**

We have successfully demonstrated plasmon measurements to evaluate charge qubits, which lays the foundation for plasmon-based cQED. The advantages of high-$Z$ resonators are expected in the strong coupling regime ($g_c > \kappa, \gamma$)[26]. Several strategies can be considered to achieve this goal.

First, both γ and κ can be reduced by optimizing device structure and material quality. The qubit decay rate γ of the present device might be raised by external electrical noise or background charge fluctuations[27,28], and thus γ can be reduced significantly, as reported in recent Ref. 16 with γ < 5 MHz. The $Q$ factor of the present resonator remains low possibly due to the nearby charge puddles (see Supplementary Fig. 4), and thus higher $Q$ is anticipated with high-quality materials, as reported in Ref. 29 with $Q$ ~ 18. We should have significant room for improving the cQED environment. Second, the coupling strength, $g_c/f_\text{r} \propto \eta\sqrt{Z}$, can be enhanced by increasing $Z$ and η. This is encouraging for entering the ultra-strong coupling regime ($g_c/f_r > 0.1$). The impedance $Z = h/(\nu e^2)$ can be increased by choosing smaller $\nu = 2, 1, 1/3, \ldots$ towards the fractional QH regimes. The coupling efficiency η ($\approx$ 0.032 in the present device) can also be increased by optimizing the capacitive coupling conditions in the present device (see Supplementary Note 6) or by introducing screening electrodes so that one dot couples more strongly to the resonator than the other, as demonstrated in some papers with large η ~ 0.3.[16,30] Figure 5 shows how the figure of merit $g_c/f_r$ increases with $Z$ and η in comparison with reported values for DQD based cQED systems (see Supplementary Note 9)[15-17,30-35]. The demonstrated scheme provides unique opportunities to explore strong plasmon–matter interactions and enable novel approaches to quantum state manipulation.

Alternatively, the high impedances of 2D topological edge channels are seen in various materials such as QH states in graphene[36-38], quantum spin Hall states in HgTe/CdTe and InAs/GaSb[39-42], and quantum anomalous Hall states in magnetically doped tetradymite semiconductors[43,44]. There might be suitable materials to integrate a plasmon resonator and a qubit in the strong coupling regime.

In addition to high-$Z$ resonators, QH resonators can provide novel cQED functionalities. For example, the extremely slow plasmon velocity[45], as compared to the light speed, is advantageous for integrating small multiple resonators on a chip. Chirality can be used for non-reciprocal coupling between multiple qubits annularly attached to a chiral plasmon resonator, which can be elaborated from a microwave circulator[46]. Furthermore, it would be intriguing to explore the coupling to topological quasiparticles like anyons in fractional QH states[47].

**Methods**

**Device Fabrication**

The device is fabricated using a GaAs/AlGaAs heterostructure with a two-dimensional electron gas (2DEG) 100 nm beneath the surface having an electron density of $1.8 \times 10^{11}$ cm$^{-2}$. The fabrication process involves several lithographic and metallization steps. Initially, the mesa structure of the device is defined by optical lithography and wet chemical etching. Then, the ohmic contacts are fabricated by using photolithography, followed by metal deposition and alloying. The fine gate electrodes in the plasmon resonator and double quantum dots (DQDs) are fabricated by using the electron-beam lithography, followed by electron-beam evaporation of Ti/Au (10/20 nm) metal films. Finally, large gate electrodes, including those for coplanar microwave transmission lines and bonding pads, are fabricated using photolithography and deposition of Ti/Au (10/250 nm) metal films.

The plasmon resonator is defined in a disk-shaped 2DEG region with a radius of ~18 μm. Two arc-shaped gate electrodes located in the left and right of the resonator are patterned with 36 μm in length and 5 μm in width, which are used as input and output ports for the plasmon signal. Adjacent to the plasmon resonator, several narrow gates with 120 nm in width are patterned for defining the DQD with each dot in a size of approximately 100×100 nm$^2$. The input and output ports of the resonator are connected to the center conductor of the coplanar microwave transmission lines, which has 32 μm in width and spaced 18 μm from the ground planes, providing a characteristic impedance of 50 Ω. The coplanar transmission lines are connected to external coaxial cables to enable the measurement of the plasmon transmission characteristics.

**Spectrum Measurement**

The transmission characteristics of the plasmon resonator (Figs. 1c, 1d, and 1e) are investigated using a vector network analyzer (VNA). The microwave signal generated from Port 1(P1) of the VNA first goes through the coaxial cables in a dilution refrigerator, and then reach the coplanar transmission line in the sample. The coplanar transmission line is connected to the input electrode ($G_{inj}$) of the resonator, which allows the microwave signal excite plasmon mode in the channel. The transmission signal through the edge channel is detected using the output electrode ($G_{det}$). The returning signal through the coaxial cables is subsequently amplified by a low-temperature amplifier (26 dB gain) at the 4 K stage in the dilution refrigerator and a room-temperature amplifier (38 dB gain) at 300 K. Finally, the amplified signal reaches Port 2(P2) of the VNA and the amplitude and phase of the signal are measured.

We employed a homodyne detection scheme (Fig. 2b) to measure the plasmon transmission spectra for qubit detection. To improve the detection signal, a lock-in technique is utilized to modulate the microwave signal at frequency $f$ ~133 Hz. In this detection scheme, the microwave signal is generated by an external microwave source instead of VNA, and the VNA detection is replaced with an IQ mixer for signal demodulation, as seen in Fig. 2b. The demodulated signal is filtered before the measurement of the in-phase ($I$) and quadrature ($Q$) components. From the measured signal, we calculate the amplitude based on $A = \sqrt{I^2 + Q^2}$ and phase based on $\phi = \arg\left(\frac{Q}{I}\right)$.

**Quantum Master Equation**

Theoretically, we model the DQD charge qubit as an effective two-level system interacting with a single plasmon mode in the resonator. The coupled system is described by the Jaynes–Cummings type model[26]

$$H = hf_r a^+ a + \frac{hf_q}{2}\sigma_z + hg(a^+\sigma_- + a\sigma_+) + H_p \quad (1)$$

where $a(a^+)$ is the plasmon annihilation (creation) operator, $\sigma_-(\sigma_+)$ the qubit lowering (rising) operator, and $\sigma_z$ the Pauli matrix. The qubit frequency is given by $hf_q = \sqrt{\varepsilon^2 + 2t_c^2}$ with the effective plasmon-qubit coupling strength $g = g_c \sin\theta_q$ accounting for the mixing of plasmon and qubit state, where $g_c$ is the bare coupling strength and $\sin\theta_q = 2t_c/hf_q$ is the qubit transition dipole moment. A microwave of frequency $f_p$ probing the transmission through the resonator can be modeled by $H_p =$ $eV_d(e^{-i2\pi f_p t}a^+ + e^{i2\pi f_p t}a)$, where $V_d$ is a measure of the drive amplitude in the resonator. This drive causes rotation of the qubit that depends on the drive amplitude.

The dissipations in the system are described with a density matrix $\rho$ through the Markovian master equation,

$$\frac{d\rho}{dt} = -\frac{i}{\hbar}[H,\rho] + \kappa(n_{pl,th}+1)D[a]\rho + \kappa n_{pl,th}D[a^+]\rho + \gamma_1(n_{ph,th}+1)D[\sigma_-]\rho$$

$$+ \gamma_1 n_{ph,th}D[\sigma_+]\rho + \frac{\gamma_\phi D[\sigma_z]\rho}{2} \quad (2)$$

with the superoperator $D[A]\rho = A\rho A^+ - \frac{\{A^+A,\ \rho\}}{2}$. Here, $\kappa$ denotes the plasmon decay rate, $\gamma_1$ is the charge relaxation rate, and $\gamma_\phi$ is the dephasing rate. The finite base temperature $T_B$ is considered with the thermal plasmon number $n_{pl,th} = \left(e^{\frac{hf_r}{k_B T_B}} - 1\right)^{-1}$ in the resonator, and thermal phonon number $n_{ph,th} = \left(e^{\frac{hf_q}{k_B T_B}} - 1\right)^{-1}$ in the qubit environment. The total qubit decoherence rate is $\gamma = \gamma_1/2 + \gamma_\phi$. In the frame rotating at the frequency of the external probe $f_p$, we obtain

$$H' = \Delta_{r-p}a^+a + \frac{\Delta_{q-p}}{2}\sigma_z + \hbar g(a^+\sigma_- + a\sigma_+) + eV_d(a^+ + a) \quad (3)$$

where $\Delta_{r-p} = hf_r - hf_p$ denotes the energy detuning between the resonator and the probe and $\Delta_{q-p} = hf_q - hf_p$ denotes the energy detuning between the qubit and the probe. To obtain the average value of the plasmon operator $\langle a \rangle = \text{Tr}(a\rho)$ and the qubit polarization $\langle \sigma_z \rangle = \text{Tr}(\sigma_z\rho)$, we derive the Bloch equations for the evolution of the field as

$$\frac{d\langle a \rangle}{dt} = -\left(\frac{\kappa}{2} + i\frac{\Delta_{r-p}}{\hbar}\right)\langle a \rangle - i\frac{eV_d}{\hbar} - ig\langle \sigma_- \rangle \quad (4)$$

$$\frac{d\langle \sigma_- \rangle}{dt} = -\left[\frac{\gamma_1}{2}(2n_{ph,th}+1) + \gamma_\phi + i\frac{\Delta_{q-p}}{\hbar}\right]\langle \sigma_- \rangle + ig\langle a\sigma_z \rangle \quad (5)$$

$$\frac{d\langle \sigma_z \rangle}{dt} = -2ig(\langle a\sigma_+ \rangle - \langle a^+\sigma_- \rangle) - \gamma_1[(2n_{ph,th}+1)\langle \sigma_z \rangle + 1] \quad (6)$$

$$\frac{d\langle a^+a \rangle}{dt} = ig(\langle a\sigma_+ \rangle - \langle a^+\sigma_- \rangle) + i\frac{eV_d}{\hbar}(\langle a \rangle - \langle a^+ \rangle) - \kappa(\langle a^+a \rangle - n_{pl,th}) \quad (7)$$

To solve these equations [Eqs. (4-7)], we approximate $\langle a\sigma_z \rangle = \langle a \rangle\langle \sigma_z \rangle$, $\langle a\sigma_+ \rangle = \langle a \rangle\langle \sigma_+ \rangle$, and $\langle a^+\sigma_- \rangle = \langle a^+ \rangle\langle \sigma_- \rangle$. These approximations provide steady state solutions,

$$\langle a \rangle = \frac{-ieV_d}{\frac{\hbar\kappa}{2} + i\Delta_{r-p} + i\hbar\chi\langle\sigma_z\rangle} \quad (8)$$

$$\langle \sigma_- \rangle = \frac{\chi}{g}\langle a \rangle\langle \sigma_z \rangle \quad (9)$$

$$\langle \sigma_z \rangle = \frac{-1}{2n_{\text{ph,th}} + 1 + \frac{4|\langle a \rangle|^2 Im(\chi)}{\gamma_1}} \quad (10)$$

$$\langle a^+ a \rangle = n_{\text{pl,th}} + \frac{2|\langle a \rangle|^2 \langle \sigma_z \rangle}{\kappa} Im(\chi) - \frac{2eV_d}{h\kappa} Im(\langle a \rangle) \quad (11)$$

where $\chi = \frac{hg^2}{\Delta_{\text{q-p}} - ih\gamma_{\text{th}}}$ is electric susceptibility associated to the decoherence $\gamma_{\text{th}} = \frac{\gamma_1}{2}(2n_{\text{ph,th}} + 1) + \gamma_\phi$ and qubit-probe energy detuning $\Delta_{\text{q-p}}$. In the dispersive regime with $|\Delta_{\text{q-p}}| \gg g, \gamma_{\text{th}}$, the electric susceptibility induces dispersive shift of the plasmon frequency[48]

$$\Delta f_r = \chi \langle \sigma_z \rangle = \frac{hg^2}{\Delta_{\text{q-p}}} \langle \sigma_z \rangle. \quad (12)$$

Equations (8-11) can be solved self-consistently. The solution of $\langle \sigma_z \rangle$ evaluates the difference between the population of the excited state and the ground state in the qubit, which decreases with increasing plasmon excitation $\langle a \rangle$ at large drive strength. With $\langle \sigma_z \rangle$, the charge population in the excited state $P_e = \langle \sigma_+ \sigma_- \rangle = \frac{1 + \langle \sigma_z \rangle}{2}$ and the ground state $P_g = \langle \sigma_- \sigma_+ \rangle = \frac{1 - \langle \sigma_z \rangle}{2}$, and polarization $P_z = \langle \sigma_z \rangle$ can be obtained. The average number of plasmons in the resonator can be calculated based on Eq. (11) and for probe frequency $f_p$ close to resonance $f_r$ ($\Delta_{\text{r-p}} = 0$), $n_{\text{pl}} = \langle a^+ a \rangle \approx n_{\text{pl,th}} + n_{\text{pl,drv}}$ with the driven plasmon number $n_{\text{pl,drv}} \approx \left(\frac{2eV_d}{h\kappa}\right)^2$, where we used the weak coupling $g \ll \kappa, \gamma_{\text{th}}$ in the approximation.

**Input–Output Theory**

To model the transmission coefficients, we use the input-output relations $\langle a_{in,\alpha} \rangle + \langle a_{out,\alpha} \rangle = \sqrt{\kappa_\alpha} \langle a \rangle$ with $\alpha = 1, 2$ as port index. When the left Port1 is used as the plasmon injection, we obtain the transmission coefficients

$$S_{21} = \frac{\langle a_{out,2} \rangle}{\langle a_{in,1} \rangle} = \frac{-i\sqrt{\kappa_1 \kappa_2}}{\frac{\kappa}{2} + i\frac{\Delta_{\text{r-p}}}{h} + i\chi \langle \sigma_z \rangle} \quad (13)$$

where we used $\langle a_{in,1} \rangle = \frac{eV_d}{\sqrt{\kappa_1}}$ and $\langle a_{in,2} \rangle = 0$. As the experimental probe frequency $f_p$ is set close to the resonance frequency $f_r$, we can use $\Delta_{\text{r-p}} = 0$ in Eq. (13), and substituting $\chi = \frac{hg^2}{\Delta_{\text{q-p}} - ih\gamma_{\text{th}}}$ yields the amplitude of the transmitted signal

$$A = |S_{21}| = \frac{\sqrt{\kappa_1\kappa_2}}{\left|\frac{\kappa}{2} - \frac{h^2 g^2 \gamma_{\text{th}}\langle\sigma_z\rangle}{\Delta_{\text{q-p}}^2 + h^2\gamma_{\text{th}}^2} + i\frac{hg^2\Delta_{\text{q-p}}\langle\sigma_z\rangle}{\Delta_{\text{q-p}}^2 + h^2\gamma_{\text{th}}^2}\right|} \quad (14)$$

and the phase

$$\phi = -\arg(S_{21}) = \tan^{-1}\left[\frac{2h^2 g^2 \Delta_{\text{q-p}}\langle\sigma_z\rangle}{h\kappa(\Delta_{\text{q-p}}^2 + h^2\gamma_{\text{th}}^2) - 2h^3 g^2\gamma_{\text{th}}\langle\sigma_z\rangle}\right]. \quad (15)$$

In the limit of weak coupling $g \ll \kappa, \gamma_{\text{th}}$, Equation (14) can be further simplified as,

$$A = |S_{21}| \approx \frac{\sqrt{\kappa_1\kappa_2}}{\sqrt{\left(\frac{\kappa}{2}\right)^2 - \kappa\frac{h^2 g^2 \gamma_{\text{th}}\langle\sigma_z\rangle}{\Delta_{\text{q-p}}^2 + h^2\gamma_{\text{th}}^2}}} \approx \frac{2\sqrt{\kappa_1\kappa_2}}{\kappa} + \frac{4\sqrt{\kappa_1\kappa_2}h^2 g^2\gamma_{\text{th}}\langle\sigma_z\rangle}{\kappa^2(\Delta_{\text{q-p}}^2 + h^2\gamma_{\text{th}}^2)} \quad (16)$$

and Eq. (15) can be simplified as

$$\phi = -\arg(S_{21}) \approx \tan^{-1}\left[\frac{2h^2 g^2 \Delta_{\text{q-p}}\langle\sigma_z\rangle}{h\kappa(\Delta_{\text{q-p}}^2 + h^2\gamma_{\text{th}}^2)}\right] \approx \frac{2h^2 g^2 \Delta_{\text{q-p}}\langle\sigma_z\rangle}{h\kappa(\Delta_{\text{q-p}}^2 + h^2\gamma_{\text{th}}^2)}, \quad (17)$$

where we used an approximation of $\tan^{-1}(x) \approx x$.

Next, we calculate the spectral shift of $A$ and $\phi$ with respect to the reference amplitude $A_0$ and phase $\phi_0$ at large qubit detuning $\varepsilon \gg 0$ (or $\Delta_{\text{q-p}} \gg 0$). By taking the limit $\Delta_{\text{q-p}} \gg 0$ for Eqs. (16) and (17), we obtain

$$A_0 = \frac{2\sqrt{\kappa_1\kappa_2}}{\kappa} \quad (18)$$

$$\phi_0 = 0 \quad (19)$$

Therefore, the spectral shift $\Delta A = A - A_0$ and $\Delta\phi = \phi - \phi_0$ can be written as

$$\Delta A = A - A_0 \approx \frac{4\sqrt{\kappa_1\kappa_2}h^2 g^2\gamma_{\text{th}}\langle\sigma_z\rangle}{\kappa^2(\Delta_{\text{q-p}}^2 + h^2\gamma_{\text{th}}^2)}, \quad (20)$$

$$\Delta\phi = \phi - \phi_0 \approx \frac{2hg^2\Delta_{\text{q-p}}\langle\sigma_z\rangle}{\kappa(\Delta_{\text{q-p}}^2 + h^2\gamma_{\text{th}}^2)}. \quad (21)$$

Equation (21) can be expressed as a function of impedance $Z$ and quality factor $Q = f_r/\kappa$ as

$$\Delta\phi \approx 2f_r(\eta e \sin\theta_{\text{qubit}})^2 \frac{\Delta_{\text{q-p}}\langle\sigma_z\rangle}{\Delta_{\text{q-p}}^2 + h^2\gamma_{\text{th}}^2}ZQ. \quad (22)$$

The experimental data in Fig. 3 in the main text and Fig. 7 in the Supplementary Note 7 are fitted by using Eq. (21), where $\langle\sigma_z\rangle$ is obtained by self-consistently solving Eqs. (8-10) with the plasmon voltage $V_d = 7.5$ μV (for $P = -87.5$ dBm). Under this drive, the qubit remains mostly in the ground state with population $P_g \cong 0.7$ and small excited state population $P_e \cong 0.3$. Fitting the data in Figs. 3(a) and 3(b) suggests the maximal coupling strength $g_c \cong 55$ MHz. In fitting the data of Figs. 3(c), $g_c$ is fixed to extract the parameters of interdot coupling $t_c$, qubit relaxation rate $\gamma_1$, dephasing rate $\gamma_\phi$. As the $\gamma_1$ and $\gamma_\phi$ cannot be determined with high accuracy, the total decoherence rate $\gamma = \gamma_1/2 + \gamma_\phi$ is studied in the main text.

The Equations (20) and (21) are also used to simulate the drive dependence of $\Delta A$ and $\Delta \phi$ shown in Figs. 4c and 4d, respectively, where the normalized values of $\frac{\Delta A}{\Delta A_0} = \frac{\langle \sigma_z \rangle}{\langle \sigma_z \rangle_0}$ and $\frac{\Delta \phi}{\Delta \phi_0} = \frac{\langle \sigma_z \rangle}{\langle \sigma_z \rangle_0}$ are compared with the experimental data. The values of $\Delta A_0$ and $\Delta \phi_0$ are calculated at zero drive limit ($V_d = 0$) with qubit polarization $\langle \sigma_z \rangle_0 = \frac{-1}{2 n_{\text{ph,th}} + 1}$.

## Data availability

The data that support the findings of this article are available via Figshare [https://doi.org/10.6084/m9.figshare.29312978].

**Acknowledgments**

We acknowledge the fruitful discussions with Yasuhiro Tokura, and Tokuro Hata. The device fabrication of this work was supported by the "Advanced Research Infrastructure for Materials and Nanotechnology in Japan (ARIM)" of the Ministry of Education, Culture, Sports, Science and Technology (MEXT). T.F. acknowledges support from the JSPS KAKENHI (Grant Numbers JP24K00552, JP24H00827). CJ.L. acknowledges support from the JSPS KAKENHI (Grant Numbers JP24K06915) and JST PRESTO (Grant Numbers JPMJPR225C).

**Author Contributions**

CJ.L. and T.F. conceived and supervised the project. T.A. and K.M. grew the heterostructure. CJ.L. fabricated the device and performed the measurements with help from K.T.. CJ.L. and T.F. analyzed the data and wrote the paper. All authors discussed the results.

**Competing Interests**

The authors declare no competing interests.

**Figure Legends/Captions**

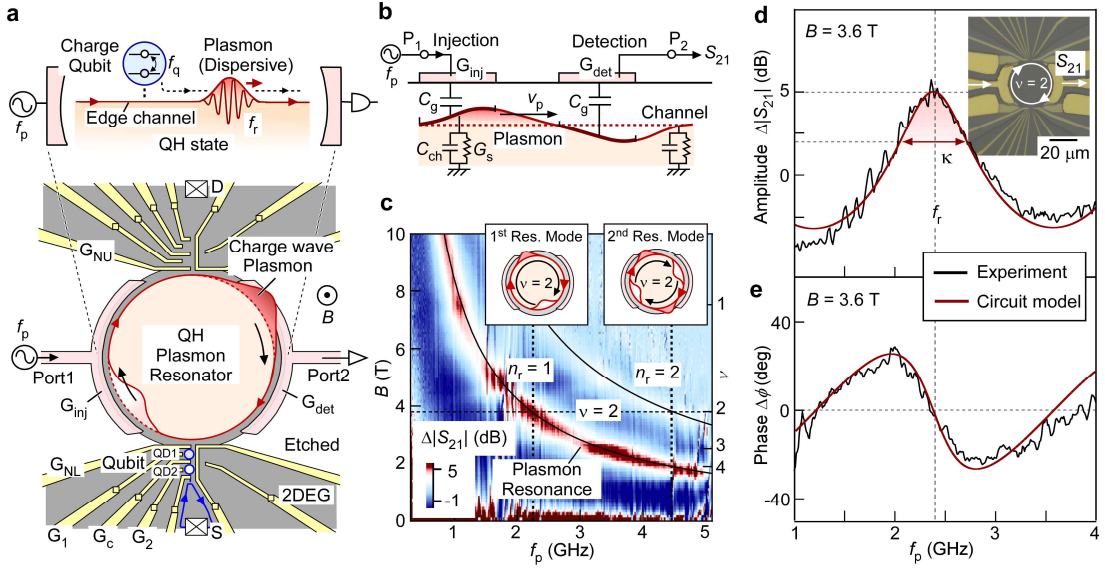

**Figure 1 | Schematic of the device and plasmon resonance characteristics. a**, Plasmon cQED system for dispersive detection of a charge qubit (upper panel). Schematic of the device (lower panel) consisting of a plasmon resonator (central region) and a DQD qubit (QD1 and QD2). The plasmon resonator is formed with circular QH channels propagating along the disk-shaped 2DEG region defined by gate electrode $G_{inj}$ and $G_{det}$. The 2DEG outside the resonator (the white regions) has been removed by wet etching. The resonant mode is represented by wave packets with excess and deficit charges. The DQD is gate-defined in the vicinity of the resonator. **b**, Circuit model for the edge channel with coupling capacitances, $C_{ch}$ and $C_g$, and dissipative conductance $G_s$. The plasmon resonance is characterized by transmission coefficient $S_{21}$ from $G_{inj}$ to $G_{det}$. (See Methods for details.) **c**, Transmission spectra $\Delta|S_{21}|(f_p, B)$ obtained at $V_{Gres}$ = -0.5 V applied on $G_{inj}$ and $G_{det}$. The right axis shows the filling factor ν. The $n_r$ = 1 and 2 plasmon resonances are seen with the corresponding charge density profiles illustrated in the insets. **d**, Amplitude $\Delta|S_{21}|$ and **e**, phase $\Delta\phi$ obtained at $B$ = 3.6 T. The red lines are calculated with a circuit model. An optical image of the device is shown in the inset to **d**.

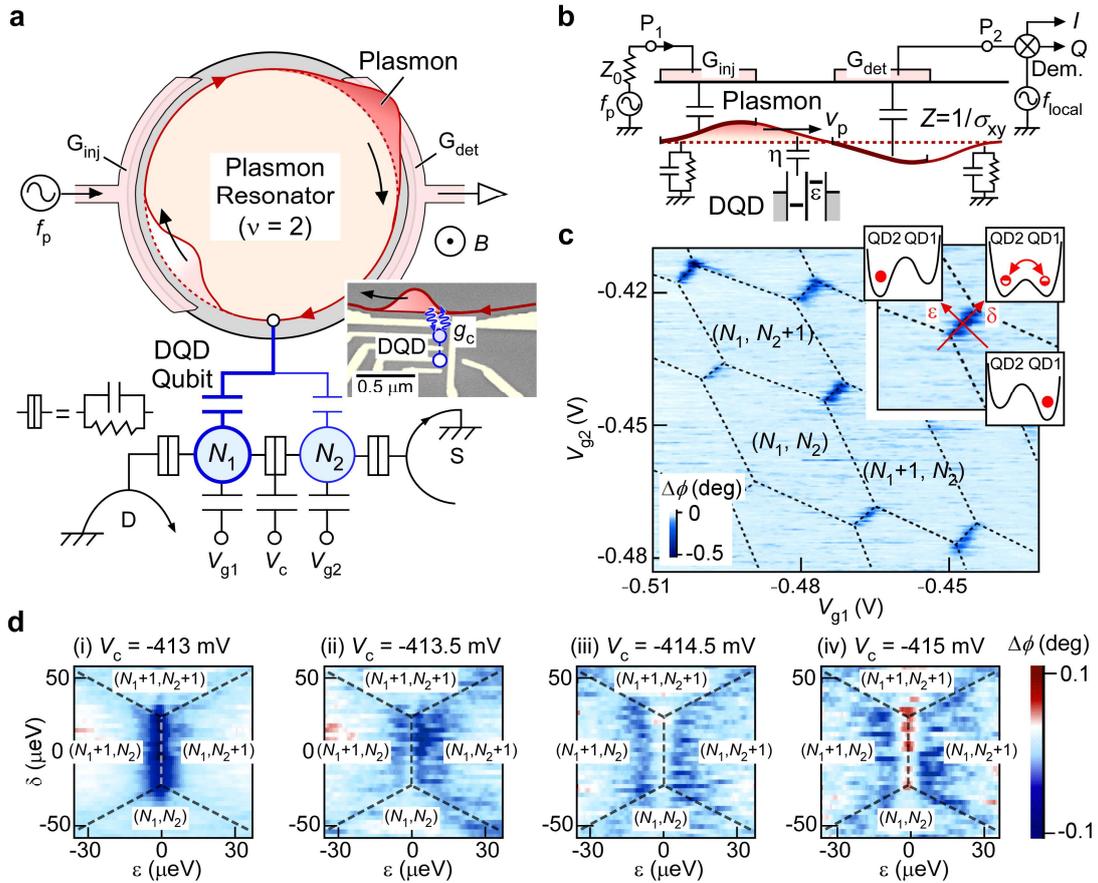

**Figure 2 | Plasmon phase spectroscopic response to DQD states. a**, Schematic of the coupled system with the qubit formed in the DQD near the plasmon resonator. The inset shows an SEM image of the DQD from a control device. Gate voltages $V_{g1}$ and $V_{g2}$ are used to control the static energy in QD1 and QD2, respectively, and the gate voltage $V_c$ is used to control the interdot tunneling coupling $t_c$. **b**, The homodyne detection scheme for extracting the amplitude and phase response of the plasmon spectra. **c**, Plasmon phase response $\Delta\phi$ plotted as a function of the voltages $V_{g1}$ and $V_{g2}$, with $V_c = -410$ mV and $B = 3.6$ T, measured at frequency $f_p = 2.5$ GHz $\cong f_r$, with an input power $P = -87.5$ dBm, where $f_r$ is the plasmon resonance frequency. The charge stability diagram of the DQD is outlined with black dashed lines with charge states labeled as $(N_1, N_2)$, where $N_1$ and $N_2$ denote the number of electrons in QD1 and QD2, respectively. **d**, $\Delta\phi$ measured around $V_{g1} = -480$ mV and $V_{g2} = -436$ mV for one particular interdot charge transfer line, under gate voltage (i) $V_c = -413$ mV, (ii) -413.5 mV, (iii) -414.5 mV, and (iv) -415 mV. The horizontal and vertical axes denote $\varepsilon$ (detuning) and $\delta$ of the DQD, which represent the difference and the average of the chemical potentials of the two dots, respectively. They are controlled by simultaneously sweeping the gate voltage $V_{g1}$ and $V_{g2}$ along the directions illustrated in the inset to **c**. Their values are converted from the lever-arm factors. The change in the phase shift from negative to positive around $\varepsilon = 0$ indicates a transition from dispersive to resonant coupling regime.

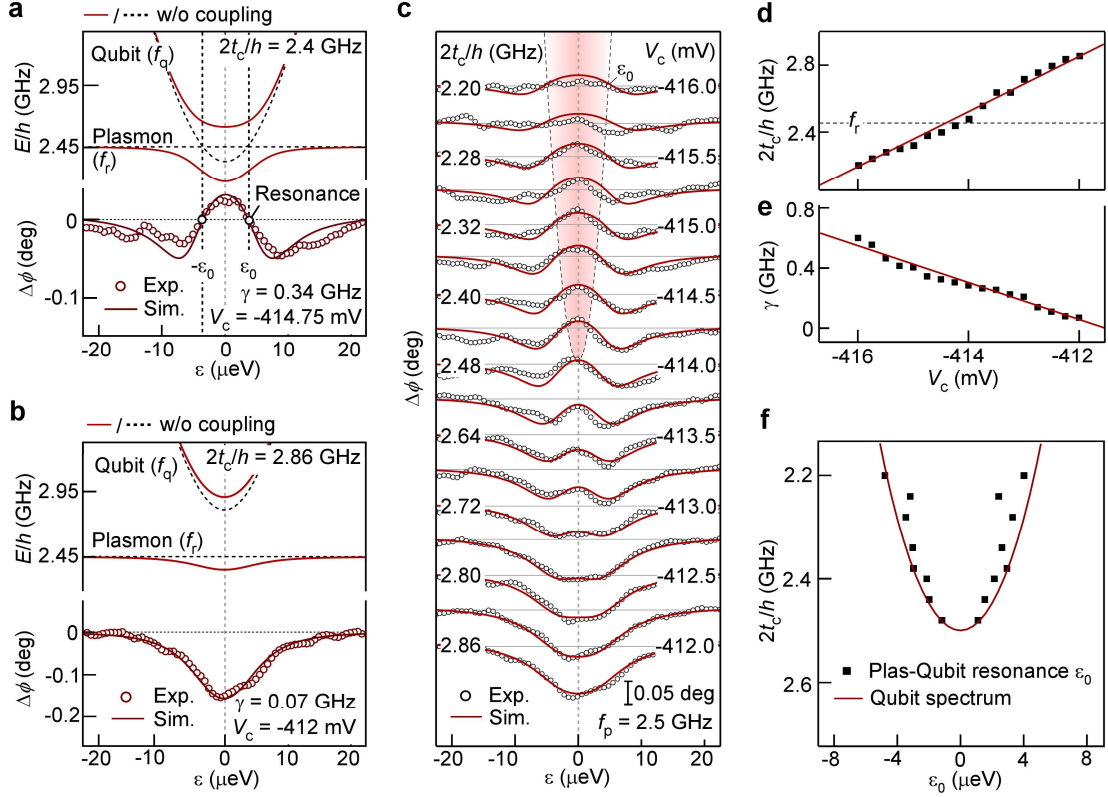

**Figure 3 | Plasmon phase spectroscopy for qubit state readout. a-b,** Phase response $\Delta\phi$ (lower panel) and energy spectra of the system (upper panel) plotted as a function of DQD detuning $\varepsilon$. $\Delta\phi$ (the red circles) is measured at frequency $f_p = 2.5$ GHz $\cong f_r$ and an input power $P = -87.5$ dBm, with gate voltage $V_c = -414.75$ mV in **a** for resonant coupling regime ($2t_c/h < f_r$), and $V_c = -412$ mV in **b** for dispersive coupling regime ($2t_c/h > f_r$). The eigenenergies of the system are calculated for the coupled case with $g_c = 55$ MHz (solid lines) and without coupling (dashed lines). The two black circles in **a** indicate the plasmon-qubit resonance at the critical detuning $\varepsilon_0$. The experimental data (circles) are fitted using theoretical simulations (solid lines) based on quantum master equations. The fitting parameters are the coherent coupling strength $g_c$, interdot tunneling coupling $t_c$ and qubit relaxation rate $\gamma_1$, and qubit dephasing rate $\gamma_\phi$. Qubit decoherence rate is defined as $\gamma = \gamma_1/2 + \gamma_\phi$. In simulation, we used the drive voltage $V_d = 7.5$ μeV for the input power $P = -87.5$ dBm. Parameters $g_c = 55 \pm 4$ MHz, $t_c/h = 1.2 \pm 0.02$ GHz, $\gamma_1 \simeq 0.24$ GHz, $\gamma_\phi \simeq 0.22$ GHz ($\gamma = 0.34 \pm 0.11$ GHz) are obtained from **a,** and $g_c = 55 \pm 4$ MHz, $t_c/h = 1.43 \pm 0.03$ GHz, $\gamma_1 \simeq 0.1$ GHz, $\gamma_\phi \simeq 0.02$ GHz ($\gamma = 0.07 \pm 0.04$ GHz) from **b**. **c,** Waterfall plot of $\Delta\phi$ as a function of $\varepsilon$ for various $V_c$ values in the range from -412 to -416 mV with a step of 0.25 mV. Solid lines show the theoretical fits obtained using the quantum master equations. **d,** $2t_c/h$ and **e,** $\gamma$ as a function of $V_c$. **f,** $2t_c/h$ plotted as a function of resonant detuning $\varepsilon_0$. The experimental data (squares) agree well with the qubit energy spectrum $2t_c = \sqrt{(hf_p)^2 - \varepsilon_0^2}$ shown as the solid line.

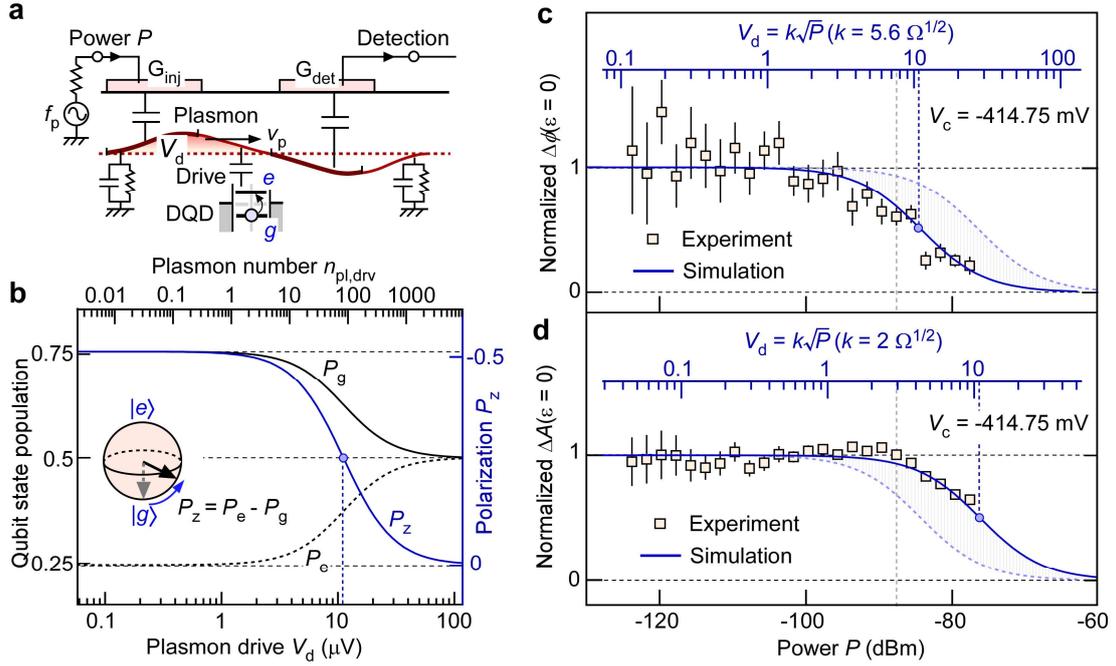

**Figure 4 | Plasmon number and qubit polarization. a**, Schematic measurement setup. The external input power $P$ excites plasmons with amplitude $V_d$, which drive the charge transition in the qubit. **b**, Simulation of qubit ground state population $P_g$, excited state population $P_e$ and polarization $P_z = P_e - P_g$ plotted as a function of plasmon drive amplitude $V_d$. $P_g$ and $P_e$ are shown on the left axis and $P_z$ is shown on the right axis. The blue circle marks $V_d$ at which $P_z$ is reduced by half. The driven plasmon number $n_{pl,drv}$ is shown on the top axis. They are calculated for the experimental condition for Fig. 3a. **c,** Normalized phase $\Delta\phi(\varepsilon = 0)$, and **d**, normalized amplitude $\Delta A(\varepsilon = 0)$ plotted as a function of external power $P$, measured at probe frequency $f_p = 2.5$ GHz $\cong$ $f_r$, and $V_c = -414.75$ mV (tunnel coupling $2t_c/h = 2.4$ GHz). The average value of $\Delta\phi(\varepsilon = 0)$ and $\Delta A(\varepsilon = 0)$ at low powers ($P < -100$ dBm) are used for normalization. The standard deviation of the mean is shown as the error bar. The simulations of $\Delta\phi(\varepsilon = 0)$ in **c** and $\Delta A(\varepsilon = 0)$ in **d** are plotted as a function of plasmon drive amplitude $V_d$ and shown as blue lines. The agreements between the experiments and simulations are achieved by scaling the horizontal axis for $V_d$. The uncertainty of the scaling factor $k$ is shown with the gray hatched region. The vertical gray dashed line in **c** and **d** marks the drive voltage $V_d$ corresponding to the power $P = -87.5$ dBm and the blue circle marks the power $P$ at which $\Delta\phi$ and $\Delta A$ is reduced by half.

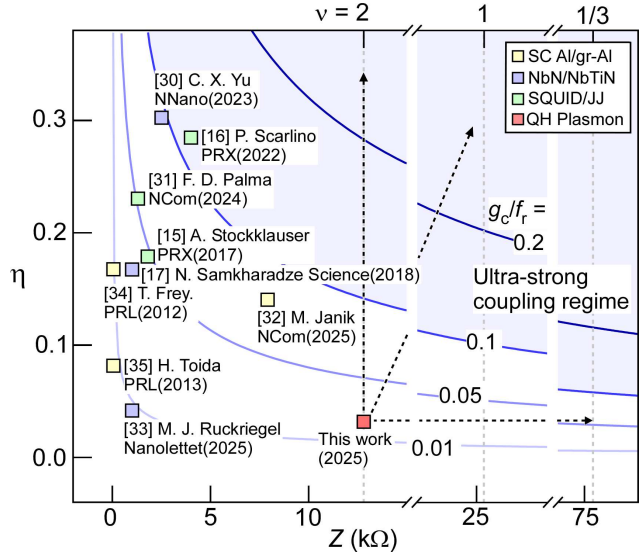

**Figure 5 | Coupling efficiency η and resonator impedance $Z$ in various DQD based cavity QED systems.** The squares show reported values of η and $Z$ for resonators made of aluminum (Al) or granular aluminum (gr-Al) superconductors (yellow squares), superconducting NbN or NbTiN materials (blue squares), high-impedance SQUID or Josephson junction (JJ) arrays (green squares), and quantum Hall plasmons of this work (the red square). The figure of merit $g_c/f_r = \eta\sqrt{\frac{Ze^2}{h}} = 0.01, 0.05, 0.1$, and $0.2$ increases with η and $Z$, as shown by the solid lines, and enters the ultra-strong coupling regime ($g_c/f_r > 0.1$) in the light blue region. The coupling strength $g_c/f_r$ of the quantum Hall cQED can be enhanced by increasing η or $Z$ (by reducing ν in the upper scale) or both, as indicated by the black arrows.

# Supplementary Information for:
# "Dispersive detection of a charge qubit with a broadband high-impedance quantum-Hall plasmon resonator"

Chaojing Lin[1,2*], Kosei Teshima[1], Takafumi Akiho[3], Koji Muraki[3], and Toshimasa Fujisawa[1]

[1]Department of Physics, Institute of Science Tokyo, 2-12-1 Ookayama, Meguro, Tokyo 152-8551, Japan.

[2]JST, PRESTO, 4-1-8 Honcho, Kawaguchi, Saitama 332-0012, Japan.

[3]Basic Research Laboratories, NTT, Inc., 3-1 Morinosato-Wakamiya, Atsugi, Kanagawa 243-0198, Japan.

*Corresponding author: Chaojing Lin (lin.c.7c98@m.isct.ac.jp)

## Supplementary Note 1: Low-Frequency Transport Characteristics

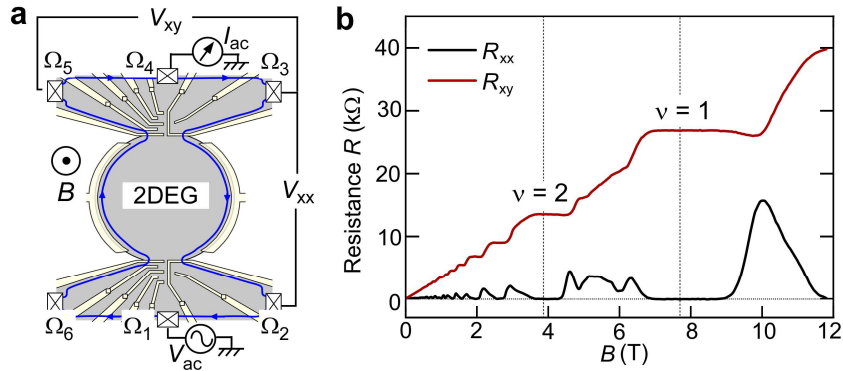

Supplementary Fig. 1. **a**, Schematic of the device and the setups for four-terminal measurements. **b**, Longitudinal resistance $R_{xx}$ and Hall resistance $R_{xy}$ plotted as a function of magnetic field $B$. The vertical dashed lines mark the QH states for $\nu = 2$ at $B = 3.8$ T, and $\nu = 1$ at $B = 7.6$ T.

We use the four terminal measurements to obtain the basic quantum Hall (QH) characteristics of the device. Supplementary Figure 1a shows the device structure and the measurement setup. Several ohmic contacts $\Omega_1$-$\Omega_6$ are used in the measurement, which are fabricated along the edge. By applying a perpendicular magnetic field $B$, the QH state is formed with a Landau level filling factor $\nu$. The field direction in out of the plane corresponds to the clockwise motion of the edge channels. The four-terminal measurements are performed by applying a source voltage $V_{ac} = 30$ μV (37 Hz) to the ohmic contact $\Omega_1$, and measuring the current $I_{ac}$ at $\Omega_4$, the longitudinal voltage $V_{xx}$ between $\Omega_2$ and $\Omega_3$, and the Hall voltage $V_{xy}$ between $\Omega_3$ and $\Omega_5$, as shown in Supplementary Fig. 1a. The calculated resistance $R_{xx} = V_{xx}/I_{ac}$ and $R_{xy} = V_{xy}/I_{ac}$ are plotted as a function of magnetic field $B$ in Supplementary Fig. 1b. QH states for $\nu = 2$ and $\nu = 1$ are identified at $B = 3.8$ T, and 7.6 T, respectively, indicated by the dashed vertical lines. The observed vanishing $R_{xx}$ and quantized $R_{xy}$ suggest that these states are well-developed with negligible bulk scattering. For qubit detection experiments, we focus on the state formed at $B = 3.6$ T with $\nu \cong 2$.

## Supplementary Note 2: Distributed Circuit Model of the Plasmon Resonator

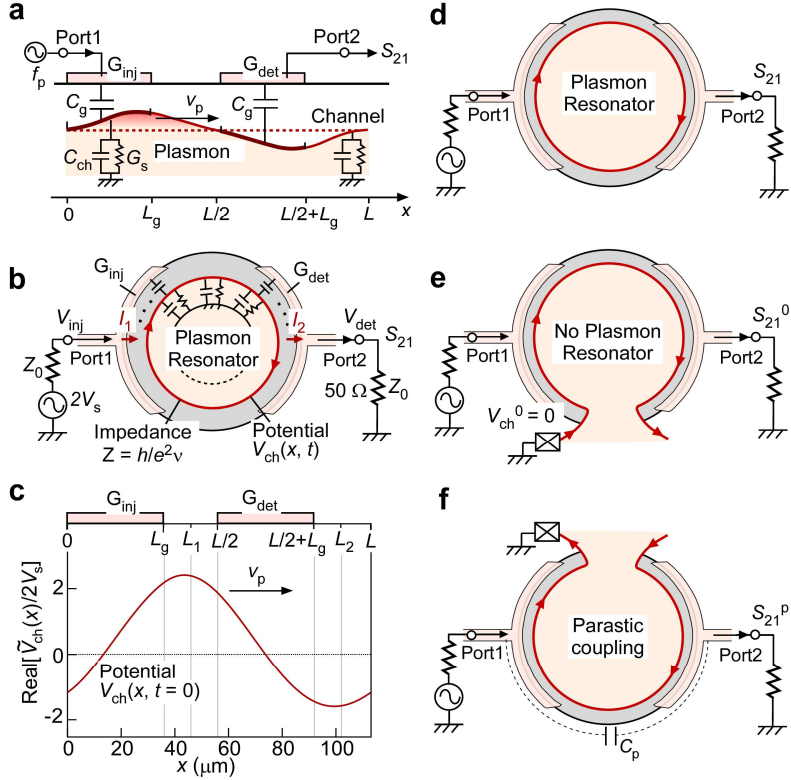

Supplementary Fig. 2. **a**, Distributed circuit model for the plasmon resonator. **b**, Measurement setup for the plasmon resonator. **c**, Real part of edge potential $\tilde{V}_{ch}(x)/2V_s$ plotted as a function of spatial axis $x$, where $2V_s$ is the source voltage shown in **b**. **d**, Ring-shaped edge channel configuration for calculation of $S_{21}$ **e**, Edge channel configuration for calculation of $S_{21}^0$. The lower constriction of the resonator is opened with input voltage $V_{ch}^0 = 0$. **f**, Edge channel configuration for calculation of $S_{21}^p$. The up constriction of the resonator is opened. The transmission is associated to a parasitic capacitance $C_p$.

To obtain a quantitative understanding of the resonance spectra, we developed a distributed circuit model for the plasmon resonator as shown in Supplementary Fig. 2a. Generally, charge density $\rho_{ch}$ and the voltage $V_{ch}$ of the edge channel are related by $\rho_{ch} = C_\Sigma V_{ch}$ with a proportional factor $C_\Sigma$ (capacitance per unit length). For convenience, $C_\Sigma$ can be decomposed into capacitances $C_g$ for the gates defining the resonator, and $C_{ch}$ for the rest including the QH bulk insulator. While a small ungated region is seen from the gate geometry shown in the Fig. 1a and optical image of the device in the insert of Fig. 1d in the main text, we assume uniform $C_\Sigma$ along the channel for simplicity. The resonator is designed symmetrically with a same length $L_g = 36$ μm for the injection gate $G_{inj}$ and the detection gate $G_{det}$ with the total resonator perimeter $L = 113$ μm.

Application of time-dependent microwave voltage $V_{inj}(t)$ to the injection gate induces a displacement current density, $C_g d(V_{inj} - V_{ch})/dt$, in the channel. This excites a plasmon wave with $V_{ch}(x, t)$. The electrical current of the edge channel is $I_{ch} = \sigma_{xy} V_{ch}$, where $\sigma_{xy}$ (= $Z^{-1}$) is the Hall conductance. The decay of plasmons in the channel is accounted for by introducing a scattering conductance $G_s$ between the channel and the ground. The current conservation law provides the wave equation in a form,

$$C_\Sigma \frac{dV_{ch}(x,t)}{dt} = -\sigma_{xy}\frac{dV_{ch}(x,t)}{dx} + C_{inj}\frac{dV_{inj}(t)}{dt} - G_s V_{ch}(x,t) \text{ at } 0 \leq x \leq L_g \quad (S1)$$

$$C_\Sigma \frac{dV_{ch}(x,t)}{dt} = -\sigma_{xy}\frac{dV_{ch}(x,t)}{dx} - G_s V_{ch}(x,t) \text{ at } L_g \leq x \leq \frac{L}{2} \quad (S2)$$

$$C_\Sigma \frac{dV_{ch}(x,t)}{dt} = -\sigma_{xy}\frac{dV_{ch}(x,t)}{dx} + C_{inj}\frac{dV_{det}(t)}{dt} - G_s V_{ch}(x,t) \text{ at } \frac{L}{2} \leq x \leq \frac{L}{2} + L_g \quad (S3)$$

$$C_\Sigma \frac{dV_{ch}(x,t)}{dt} = -\sigma_{xy}\frac{dV_{ch}(x,t)}{dx} - G_s V_{ch}(x,t) \text{ at } \frac{L}{2} + L_g \leq x \leq L \quad (S4)$$

We solve these equations by using periodic boundary conditions at $x = 0, L_g, L/2, L/2+L_g$. For an excitation voltage of the form, $V_{inj}(t) = \tilde{V}_{inj} e^{i\omega t}$ with complex amplitude $\tilde{V}_{inj}$, on $G_{inj}$, the complex amplitude of the potential of the channel is given by $V_{ch}(x,t) = \tilde{V}_{ch}(x) e^{i\omega t}$ with $\tilde{V}_{ch}(x)$ given by

$$\tilde{V}_{ch}(x) = \frac{i\omega C_g}{G_s + i\omega C_\Sigma}\left\{\left[1 - \frac{1-e^{-ik(L-L_g)}}{1-e^{-ik}}e^{-ikx}\right]\tilde{V}_{inj} + \frac{e^{-ik(\frac{L}{2}-L_g)} - e^{-ik(\frac{L}{2})}}{1-e^{-ikL}}e^{-ikx}\tilde{V}_{det}\right\}$$

at $0 \leq x < L_g$ (S5)

$$\tilde{V}_{ch}(x) = \frac{i\omega C_g}{G_s + i\omega C_\Sigma}\left(\frac{1-e^{-ikL_g}}{1-e^{-ikL}}\right)\left[\tilde{V}_{inj} + e^{-ik(\frac{L}{2})}\tilde{V}_{det}\right]e^{-ik(x-L_g)} \text{ at } L_g \leq x < \frac{L}{2} \quad (S6)$$

$$\tilde{V}_{ch}(x) = \frac{i\omega C_g}{G_s + i\omega C_\Sigma}\left\{\left[1 - \frac{1-e^{-i(L-L_g)}}{1-e^{-ikL}}e^{-ik(x-\frac{L}{2})}\right]\tilde{V}_{det} + \frac{e^{ikL_g}-1}{1-e^{-ikL}}e^{-ikx}\tilde{V}_{inj}\right\} \text{ at } \frac{L}{2} \leq x$$

$< \frac{L}{2} + L_g$ (S7)

$$\tilde{V}_{ch}(x) = \frac{i\omega C_g}{G_s + i\omega C_\Sigma}\left(\frac{1-e^{-ikL_g}}{1-e^{-ikL}}\right)\left[\tilde{V}_{inj} + e^{ik(\frac{L}{2})}\tilde{V}_{det}\right]e^{-ik(x-L_g)} \text{ at } \frac{L}{2} + L_g \leq x < L \quad (S8)$$

where $k = \omega/v_p + iG_s/\sigma_{xy}$ is the complex wave number with $v_p = \sigma_{xy}/C_\Sigma$ the plasmon velocity and $G_s/\sigma_{xy}$ the plasmon decay rate. The factor $\frac{1}{1-e^{-ikL}}$ in Eqs. (S5-S8) determines the resonant frequency $\omega_r = 2\pi\sigma_{xy}/LC_\Sigma$.

For the measurement circuit shown in Supplementary Fig. 2b, the voltage and current at injection and detection ports can be related by the forms, $\tilde{V}_{inj} = 2\tilde{V}_s - \tilde{I}_1 Z_0$ and $\tilde{V}_{det} = \tilde{I}_2 Z_0$, where $\tilde{I}_1 = \int_0^{L_g} i\omega C_g(\tilde{V}_{inj} - \tilde{V}_{ch}) dx$ is the complex amplitude of the displacement current input through the injection gate $G_{inj}$ and $\tilde{I}_2 = \int_{\frac{L}{2}}^{\frac{L}{2}+L_g} i\omega C_g(\tilde{V}_{ch} - \tilde{V}_{det}) dx$ is the displacement current flowing out through the detection gate $G_{det}$. Here, $Z_0 = 50\ \Omega$ is the impedance in the measurement circuits.

Using the current-voltage relation and the Eq. (S5-S8), we can solve the voltage in the channel. Supplementary Figure 2c shows the spatial part of $V_{ch}(x, t = 0) = \tilde{V}_{ch}(x)$ for the first resonant mode with $C_\Sigma = C_g = 0.28$ nF/m and $G_s = 0.55$ s/m. When considering the temporal component, this voltage profile corresponds to a chiral wave propagating along the positive $x$ with a velocity $v_p$. The DQD coupled to this chiral wave is placed at $x = L_2 = 103$ μm as shown in Fig. 1a.

The transmission signal $S_{21}$ from port 1 to port 2 can be calculated as

$$S_{21}(\omega) = \frac{\tilde{V}_{det}}{\tilde{V}_s} \approx \frac{-2iZ_0(\omega C_g)^2}{G_s + i\omega C_\Sigma} \frac{(1 - e^{-ikL_g})(1 - e^{ikL_g})e^{-\frac{ikL}{2}}}{k(1 - e^{-ikL})} \quad (S9)$$

where we used $\sigma_{xy} Z_0 \ll 1$ considering that the plasmon impedance $Z = 1/\sigma_{xy}$ is much larger than the impedance $Z_0 = 50\ \Omega$ in the measurement circuits.

In analogy to the above calculation of $S_{21}$ for ring-shaped channel as shown in Supplementary Fig. 1c, we conducted similar calculation of $S_{21}^0$ for the open channel as shown in Supplementary Fig. 1d, which is realized by opening the lower constriction of the resonator by setting the gate voltage on $G_{NL}$ to zero [Fig. 1a]. As in this case the edge channel is connected to a grounded ohmic contact, we use a boundary condition $V_{ch}^0 = V_{ch}(x = 0) = 0$ in the calculation and obtain

$$S_{21}^0(\omega) = \frac{\tilde{V}_{det}}{\tilde{V}_s} \approx \frac{-2iZ_0(\omega C_g)^2}{G_s + i\omega C_\Sigma} \frac{(1 - e^{ikL_g})(1 - e^{-ikL_g})e^{-\frac{ikL}{2}}}{k} \quad (S10)$$

where we used $\sigma_{xy} Z_0 \ll 1$ to obtain the second expression.

The parasitic capacitive coupling between the input and output ports is investigated with the setup shown in Supplementary Fig. 1f, where the upper constriction of the resonator is opened by setting the gate voltage on $G_{NU}$ to zero [Fig. 1a]. This allows the plasmons to propagate to a grounded ohmic contact. The transmission signal $S_{21}^p$ is thus dominated by the parasitic capacitance $C_p$, and this can be calculated as,

$$S_{21}^p(\omega) = \frac{\tilde{V}_{det}}{\tilde{V}_s} = \frac{2i\omega Z_0 C_p}{1 + 2i\omega Z_0 C_p}. \quad (S11)$$

In the following, the experimental results for different setups are analyzed with these circuit models to characterize the fundamental properties of the plasmon resonator.

**Supplementary Note 3: Transport Characteristics of the Plasmon Resonator**

Supplementary Figure 3a shows the measured transmission spectra $|S_{21}|$, $|S_{21}^0|$ and $|S_{21}^p|$ at $B = 3.6$ T, with the measurement setups shown in Supplementary Figs. 2d- 2f. $|S_{21}|$ shows a broad resonance peak with short-period oscillations possibly originated from the external measurement circuits. By subtracting $|S_{21}|$ with the reference signal $|S_{21}^0|$, these oscillations can be removed. The subtracted $\Delta|S_{21}| = |S_{21}/S_{21}^0|$ (in a decibel unit) is presented in Supplementary Fig. 3b, showing clear resonance at frequency $f_r = 2.45$ GHz with a linewidth $\kappa = 0.6$ GHz. The formation of resonator enhances the transmission signal of ~5 dB at resonance. The overall spectrum can be fitted well by using Eqs. (S9–S10) in the Supplementary Note 2. From the fitting, we obtain two values: $C_g$ = 0.28 nF/m and $G_s$ = 0.55 S/m. The first value describes the plasmon transmission efficiency, and the second value associated with the peak width describes the plasmon decay rate. The spectrum and fitting curves are replotted in Fig. 1d of the main text.

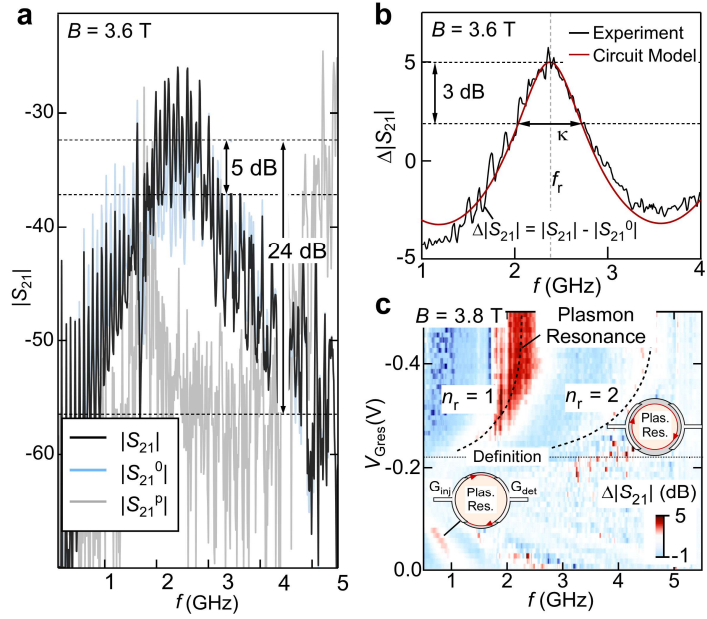

Supplementary Fig. 3. **a**, Measured transmission spectra $S_{21}$, $S_{21}^0$ and $S_{21}^p$ at $B$ = 3.6 T, with $S_{21}$ for resonator, $S_{21}^0$ for opening the lower constriction of the resonator, and $S_{21}^p$ for opening the upper constriction of the resonator. **b**, $\Delta|S_{21}| = |S_{21}/S_{21}^0|$ plotted in a decibel unit. The fitting with the circuit model is shown as the red line. **c**, $\Delta|S_{21}|$ at various gate voltage $V_{Gres}$ applied on gate $G_{inj}$ and $G_{det}$, measured at $B$ = 3.8 T. Plasmon resonance is observed for $V_{Gres}$ below the definition voltage -0.22 V, and around 0 V, with the edge channel configuration illustrated in the insert.

For comparison, Supplementary Fig. 3a also shows the measured $|S_{21}^p|$, characterizing parasitic coupling between the input and output ports. Comparing to the resonance peak, $|S_{21}^p|$ is reduced by ~24 dB and shows small transmission amplitude $|S_{21}^p|$ ~ -56 dB. This corresponds to a parasitic capacitance $C_p \cong$ 1 fF by using Eq. (S11), which is an order of magnitude smaller than the channel capacitance $C_g L_g \cong$ 10.1 fF for transmission of plasmon signal. Therefore, this small $C_p$ was neglected in the resonator model in Supplementary Note 2.

Supplementary Fig. 3c demonstrates the gate voltage tuning of the resonant spectra. Here, the gate voltage $V_{Gres}$ applied on both $G_{inj}$ and $G_{det}$ is varied. Blueshift of the resonance frequency is observed as $V_{Gres}$ is reduced below the definition voltage ($V_{Gres}$ < -0.22 V). The frequency shift can be understood as the decrease of channel capacitance $C_\Sigma$ while the channel moves away from the gate electrode. At around $V_{Gres} = 0$ V, a lower-frequency resonance is also observed, where the edge channel is formed beneath the gate $G_{inj}$ and $G_{det}$ and has a large channel capacitance. For the qubit detection, we use the resonator defined at $V_{Gres} = -0.5$ V to obtain dispersive coupling to the qubit.

**Supplementary Note 4: Field Dependence of the Quality Factor**

From the field dependence of the resonance spectra in Fig. 1c, we extract the resonance frequency $f_r$, bandwidth $\kappa$, and quality factor $Q = f_r/\kappa$ and plot the data as a function of magnetic field $B$ in Supplementary Figs. 4b–4d, together, we plot the longitudinal resistance $R_{xx}$ in Supplementary Fig. 4a for comparison. As the magnetic field $B$ increases, the resonance frequency $f_r$ decreases, reflecting the reduction of plasmon velocity. Whereas the overall bandwidth $\kappa$ becomes broader with decreasing the field, pronounced minimum develops at integer QH states at $\nu = 1, 2$, and 3. Notably, the $Q$ values at these states are nearly identical of ~4 as shown in Supplementary Fig. 4d. While the $Q$ value sensitively changes with the magnetic field, the $R_{xx}$ value remains zero over a wide field range. These observations demonstrate the high-frequency transport characteristics of plasmons and their correlation to the QH states.

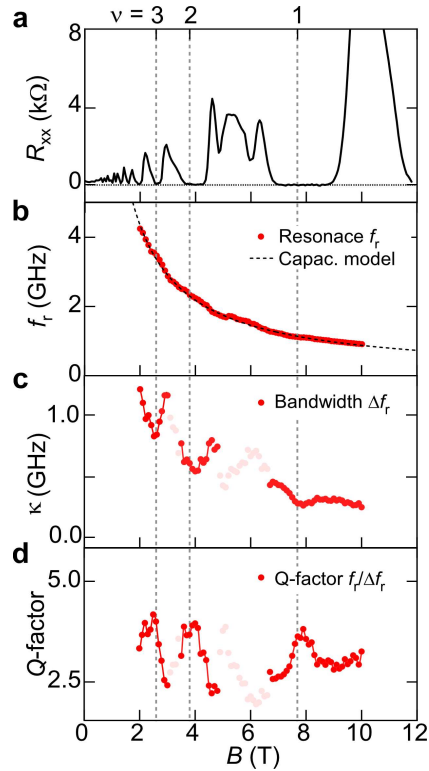

Supplementary Fig. 4. **a**, Longitudinal resistance $R_{xx}$, **b**, Resonance frequency $f_r$, **c**, Bandwidth $\kappa = \Delta f_r$, and **d**, Quality factor $Q = f_r/\Delta f_r$ plotted as a function of magnetic field $B$. $f_r$ and $\kappa$ are extracted from the magnetic field dependence of the resonance spectra in Fig. 1c in the main text.

**Supplementary Note 5: Transport Characteristics of the DQD**

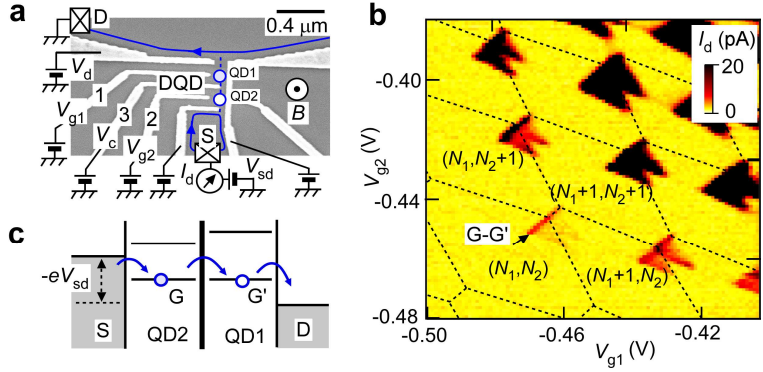

Supplementary Fig. 5. **a**, An SEM image of a control device and the measurement setup for the DQD. **b**, Tunnel current $I_d$ as a function of gate voltage $V_{g1}$ and $V_{g2}$. **c**, Transport through the DQD with energy levels in the two dots aligned in the biased window.

Supplementary Figure 5a shows an SEM image of a double quantum dot (DQD) device in the vicinity of the plasmon resonator. Five finger gates are used to define the DQD marked by two blue circles. The tunneling current $I_d$ through the DQD is measured with a dc bias voltage $V_{sd}$ (= 400 μV) applied between the source (S) and the drain (D). The energy levels in the QD1 (QD2) are tuned by applying the gate voltage $V_{g1}$ ($V_{g2}$) to gate 1 (2). The interdot tunneling coupling can be tuned by applying the gate voltage $V_c$ to gate 3.

Supplementary Figure 5b shows the stability diagram of the DQD measured at $B$ = 2 T (ν ~ 4) in the absence of microwave irradiation ($P$ = 0). The measured $I_d$ is plotted in the color scale as a function of the gate voltages $V_{g1}$ and $V_{g2}$ at $V_c$ = -0.45 V. Triangular conductive regions are located at the corners of honeycomb-shaped stability diagram represented by the dashed lines. The size of the triangles measures the transport window of $eV_{sd}$ = 400 μeV, from which the lever arm factor was determined. The spacings between triangles measures the Coulomb charging energy of ~0.7 meV for both dots and the interdot electrostatic coupling energy of ~74 μeV. Some triangles show fine peaks associated with the resonant tunneling between the ground state G-G' of the two dots. The charge transfer line along G-G' is elongated due to the finite bias, as illustrated in Supplementary Fig. 5c.

**Supplementary Note 6: Plasmon Detection of the Charge State in DQD**

In Supplementary Fig. 6, the phase response $\Delta\phi$ is presented as a function of gate voltages $V_{g1}$ and $V_{g2}$ over a wide tuning range. The charge stability diagram for electron occupations ($N_1$, $N_2$) is marked by the black dashed lines, where $N_1$ and $N_2$ denote the electron numbers in QD1 and QD2, respectively. Negative phase shifts are observed along various charge transition lines, which include the tunneling between the dots and the leads as well as interdot tunneling. The presence of finite signals indicates these transitions having a tunneling rate close to the resonance frequency.

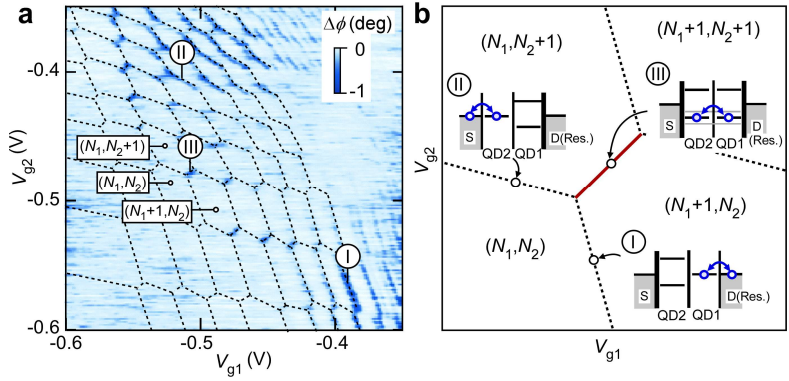

Supplementary Fig. 6. **a**, Color plot of the phase response $\Delta\phi$ as a function of gate voltages $V_{g1}$ and $V_{g2}$. The charge stability diagram is outlined by the black dashed lines. **b,** Energy diagrams for electron transport in the DQD.

Specifically, clear charging lines of QD1 appear in the lower-right region of the diagram with small negative $V_{g1}$ and large negative $V_{g2}$, (i.e., position I, $V_{g1}$ = -0.39 V, $V_{g2}$ = -0.56 V), where the charge tunnels between the QD1 and the drain (D) while QD2 is in the Coulomb blockade regime, as illustrated in Supplementary Fig. 6b. The signal for charging lines of QD2 appear in the upper-left region with small negative $V_{g2}$ and large negative $V_{g1}$, (i.e., position II, $V_{g1}$ = -0.51 V, $V_{g2}$ = -0.4 V). Interdot charge transfer between QD1 and QD2 is observed at the lower-left region (i.e., position III, $V_{g1}$ = -0.51 V, $V_{g2}$ = -0.48 V) with large negative $V_{g1}$ and $V_{g2}$. In this region, charging lines of QD1 and QD2 are not observed as the low tunneling rate. The disappearance of the signal in the upper-right region may be attributed to poorly developed discrete levels in the DQD.

The present experiment mainly focused on the configuration where the electron is strongly confined in the DQD having weak coupling to source and drain (i.e., position III in the lower-left region of the charge diagram). Fine measurements at this region are presented in Fig. 2c in the main text.

**Supplementary Note 7: Plasmon Phase Response Under Various Probe Frequency $f_p$**

The plasmon resonator has a broad bandwidth of ~ 0.6 GHz, which enables wide range detection of qubit spectra by varying the probe frequency $f_p$. Supplementary Figure 7a presents the phase $\Delta\phi$ as a function of detuning ε measured at different $f_p$ with $V_c$ = -414.75 mV. We observe results analogous to those obtained when tuning $V_c$ or $t_c$, as seen in Fig. 3c in the main text. At low $f_p$, the phase displays a single negative peak near zero detuning (ε = 0). This negative peak bifurcates as $f_p$ increases. At high $f_p$, a positive peak emerges at zero detuning with negative peaks on both sides. This transition indicates a shift of the system's coupling regime, which transitions from the dispersive coupling regime ($f_p < 2t_c/h$) to the resonant coupling regime ($f_p > 2t_c/h$).

The spectra at off-resonance can be fitted using the quantum master equation theory by assuming the plasmon frequency $f_r$ follows the probe frequency $f_p$, $f_r = f_p$. The extracted $t_c$ and $\gamma$ values are plotted in Supplementary Fig. 7b, and Fig. 7c, respectively. As we fixed the DQD configuration, $2t_c/h$ (~2.4 GHz) shows barely change, although a small increase of $\gamma$ is observed at large $f_p$. The variation of $\gamma$ (0.14-0.44 GHz) is smaller than that in the $V_c$ dependency of Fig. 3e. The small increase of $\gamma$ may originate from unwanted variations, such as the transmission power and the electrical noises in the measurement circuit. The resonance between plasmon and qubit is reached at zero-phase-shift condition ($\Delta\phi = 0$) for $f_p > 2t_c/h$. The extracted resonance detuning $\varepsilon_0$ is plotted against $f_p$ in Supplementary Fig. 7d, which agrees well with the qubit energy spectrum $f_q(\varepsilon) = \frac{\sqrt{\varepsilon^2 + 4t_c^2}}{h}$, as shown by the solid line.

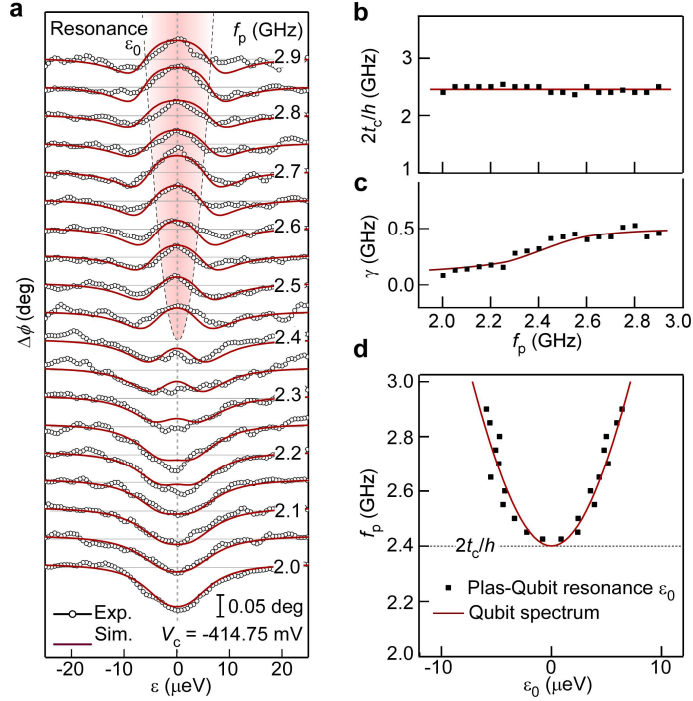

Supplementary Fig. 7. **a**, Waterfall plot of $\Delta\phi$ as a function of $\varepsilon$ for various probe frequency $f_p$ from 2 GHz to 2.9 GHz with step 0.5 GHz. Fitting curves based on the master equation theory are presented as solid lines. **b**, Extracted interdot coupling $2t_c/h$, and **c**, qubit decoherence rate $\gamma$, plotted as a function of $f_p$. **d**, Extracted plasmon-qubit resonance detuning $\varepsilon_0$ at various $f_p$ in **a**. The qubit energy spectrum is shown as the red line.

## Supplementary Note 8: Power Dependence of the Plasmon Transmission Spectra

In the resonant coupling regime, we have systematically measured plasmon transmission spectra under various input power $P$. Supplementary Figure 8a shows the power dependence of the phase response $\Delta\phi$, and Supplementary Figure 8c shows the amplitude response $\Delta A$, both plotted as a function of energy detuning $\varepsilon$. These measurements were performed at probe frequency $f_p = 2.5$ GHz $\cong f_r$ and $V_c = -414.75$ mV with $2t_c/h = 2.4$ GHz.

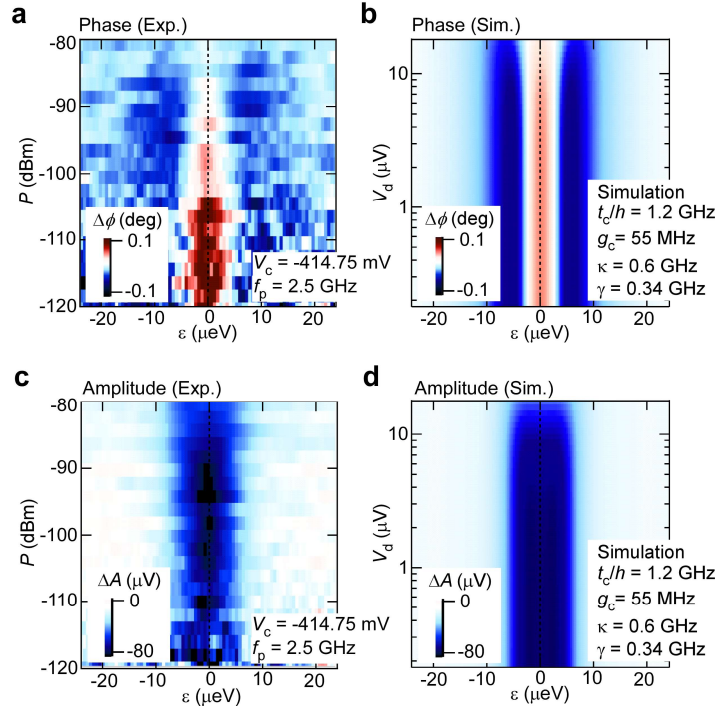

Supplementary Fig. 8. **a**, Power $P$ dependence of $\Delta\phi(\varepsilon)$ obtained at $f_p = 2.5$ GHz $\cong f_r$ and $V_c = -414.75$ mV (corresponding to $2t_c/h = 2.4$ GHz). **b**, Simulation of $\Delta\phi$ for various plasmon drive amplitude $V_d$. **c**, $P$ dependence of $\Delta A(\varepsilon)$. **d**, Simulation of $\Delta A(\varepsilon)$ for various $V_d$.

The amplitude response exhibits a single negative dip near zero detuning $\varepsilon = 0$, contrasting with the alternating negative - positive variations in the phase signal. Both $\Delta\phi$ and $\Delta A$ decrease at high power. These characteristics are quantitatively reproduced by simulations with varying plasmon drive amplitude $V_d$, as shown in Supplementary Figs. 8b and 8d. The simulations are performed based on Eqs. (20-21) in the Methods section of the main text. The polarization $\langle\sigma_z\rangle$ in these equations is obtained by self-consistently solving Eqs. (8-11) for a given drive voltage $V_d$, using the parameters of $2t_c/h = 2.4$ GHz, $g_c = 55$ MHz, $\gamma = 0.34$ GHz, and $\kappa = 0.6$ GHz. In Fig. 4 of the main text, the scaling factor $k$ in the relation of $V_d = k\sqrt{P}$ is estimated to evaluate the plasmon number in the resonator, and the charge polarization in the qubit.

**Supplementary Note 9: Coupling Strength in Various Quantum Dot Cavity QED Systems**

Supplementary Table 1 summarizes cavity and qubit parameters and their geometric factor η, and coupling strength $g_c$ in various DQD cavity QED systems. The extracted coupling strength in our system is comparable to that found in early GaAs QED systems [1-2]. As the small η ~ 0.032 in present device is the main constraint limiting the coupling strength $g_c$, further optimization of the device structure is required. For example, by using screening electrode for asymmetric resonator-qubit coupling, a large η ~ 0.3 has been reported [4,5], indicating that there is still substantial room for improvement in our system.

Supplementary Table. 1| Extracted cavity, qubit, and coupling parameters in various DQD based cavity QED systems

| Ref | Cavity | | | | | Qubit | | | Coupling | |
|---|---|---|---|---|---|---|---|---|---|---|
| | Resonator | $Z$ ($\Omega$) | $f_r$ (GHz) | $\kappa$ (MHz) | $Q$ | DQD | $\gamma$ (MHz) | $\eta$ | $g_c$ (MHz) | $g_c/f_r$ |
| [1] | Al CPW | 50 | 6.755 | 2.6 | 2630 | GaAs | 950 | 0.17 | 50 | 0.0074 |
| [2] | Al CPW | 50 | 8.3267 | 8 | 1040 | GaAs | 25 | 0.08 | 30 | 0.0036 |
| [3] | SQUID | 1.8 k | 5.02 | 12 | 417 | GaAs | 40 | 0.18 | 238 | 0.047 |
| [4] | JJ array | 4 k | 5.6 | 23 | 243 | GaAs | <5 | 0.28 | 630 | 0.11 |
| [5] | NbN | 2.5 k | 5.43 | 14 | 388 | Si NW | 9.9 | 0.3 | 513 | 0.094 |
| [6] | SQUIDs | 1.3 k | 5.01 | 63 | 80 | Ge/SiGe | 192 | 0.23 | 260 | 0.052 |
| [7] | gr-Al | 7.9 k | 7.262 | 17 | 427 | Ge/SiGe | 297 | 0.14 | 566 | 0.078 |
| [8] | NbTiN | 1 k | 6.051 | 2.7 | 2241 | Si/SiGe | 52 | 0.17 | 200 | 0.033 |
| [9] | NbTiN | 1 k | 6.033 | 8.8 | 686 | Graphene | 643 | 0.042 | 49.7 | 0.008 |
| [10] | QH Plas. | 12.8 k | 2.45 | 600 | 4 | GaAs | 150-800 | 0.032 | 55 | 0.022 |

**Supplementary References**